\pdfoutput=1

\documentclass[aps,prl,twocolumn,bibnotes]{revtex4-1}

\usepackage[breaklinks=true,backref=false,colorlinks=true]{hyperref}

\usepackage{newtxtext}
\usepackage[varvw]{newtxmath}
\usepackage[activate={true},final,tracking=true,kerning=true,spacing=true,factor=1100,stretch=20,shrink=20]{microtype}
\UseMicrotypeSet[protrusion]{basictext}
\microtypecontext{spacing=nonfrench}
\usepackage{amsmath}
\usepackage{amssymb}
\usepackage{cancel}

\usepackage{lineno}
\newcommand*\patchAmsMathEnvironmentForLineno[1]{%
   \expandafter\let\csname old#1\expandafter\endcsname\csname #1\endcsname
   \expandafter\let\csname oldend#1\expandafter\endcsname\csname end#1\endcsname
   \renewenvironment{#1}%
      {\linenomath\csname old#1\endcsname}%
      {\csname oldend#1\endcsname\endlinenomath}}% 
\newcommand*\patchBothAmsMathEnvironmentsForLineno[1]{%
   \patchAmsMathEnvironmentForLineno{#1}
   \patchAmsMathEnvironmentForLineno{#1*}}%
\AtBeginDocument{%
\patchBothAmsMathEnvironmentsForLineno{equation}%
\patchBothAmsMathEnvironmentsForLineno{align}%
\patchBothAmsMathEnvironmentsForLineno{flalign}%
\patchBothAmsMathEnvironmentsForLineno{alignat}%
\patchBothAmsMathEnvironmentsForLineno{gather}%
\patchBothAmsMathEnvironmentsForLineno{multline}% 
}

\usepackage{color}

\usepackage{graphicx}% Include figure files
\usepackage{xcolor}  % Coloured text etc.
\usepackage{transparent}
\usepackage{dcolumn}% Align table columns on decimal point
\usepackage{bm}% bold math
\usepackage{enumerate}
\usepackage{slashed}
\usepackage{simplewick}
\usepackage{comment}
\usepackage{soul}

\usepackage[english]{babel}
\usepackage{appendix}
\usepackage[utf8]{inputenc}

\usepackage{marginnote,letltxmacro}

\LetLtxMacro{\mn}{\marginnote}
\usepackage[normalem]{ulem}
\usepackage{subfigure}

\newcommand{\uu}{{\rm{u}}}
\newcommand{\bz}{\bar z}

\newcommand{\ii}{{\rm{i}}}
\newcommand{\dd}{{\rm{d}}}
\newcommand{\vv}{{\rm {v}}}

\newcommand{\nn}{\nonumber}

\def\sign{\mbox{sign}}
\newcommand{\eq}[1]{(\ref{#1})}

\newcommand{\la}{\label}
\newcommand{\ba}{\begin{align}}
\newcommand{\ee}{\end{equation}}
\newcommand{\be}{\begin{equation}}

\def\12{\frac{1}{2}}
\newcommand{\p}{\partial}
\newcommand{\en}{\end{align}}

\usepackage{color}

\def\XXint#1#2#3{{\setbox0=\hbox{$#1{#2#3}{\int}$}
     \vcenter{\hbox{$#2#3$}}\kern-.5\wd0}}

\usepackage{xparse}  %%define \restr{}{} command
\DeclareDocumentCommand{\restr}{m m o}  
{%
	\IfNoValueTF{#3}
	{
		\left.\kern-\nulldelimiterspace
		#1 
		\vphantom{\big|} 
		\right|_{#2}
	}{
		\left.\kern-\nulldelimiterspace 
		#1 
		\vphantom{\big|} 
		\right|_{#2}^{#3} }
}

\makeatletter
\let\save@mathaccent\mathaccent
\newcommand*\if@single[3]{%
  \setbox0\hbox{${\mathaccent"0362{#1}}^H$}%
  \setbox2\hbox{${\mathaccent"0362{\kern0pt#1}}^H$}%
  \ifdim\ht0=\ht2 #3\else #2\fi
  }
\newcommand*\rel@kern[1]{\kern#1\dimexpr\macc@kerna}
\newcommand*\wbar[1]{\@ifnextchar^{{\wide@bar{#1}{0}}}{\wide@bar{#1}{1}}}
\newcommand*\wide@bar[2]{\if@single{#1}{\wide@bar@{#1}{#2}{1}}{\wide@bar@{#1}{#2}{2}}}
\newcommand*\wide@bar@[3]{%
  \begingroup
  \def\mathaccent##1##2{%
    \let\mathaccent\save@mathaccent
    \if#32 \let\macc@nucleus\first@char \fi
    \setbox\z@\hbox{$\macc@style{\macc@nucleus}_{}$}%
    \setbox\tw@\hbox{$\macc@style{\macc@nucleus}{}_{}$}%
    \dimen@\wd\tw@
    \advance\dimen@-\wd\z@
    \divide\dimen@ 3
    \@tempdima\wd\tw@
    \advance\@tempdima-\scriptspace
    \divide\@tempdima 10
    \advance\dimen@-\@tempdima
    \ifdim\dimen@>\z@ \dimen@0pt\fi
    \rel@kern{0.6}\kern-\dimen@
    \if#31
      \overline{\rel@kern{-0.6}\kern\dimen@\macc@nucleus\rel@kern{0.4}\kern\dimen@}%
      \advance\dimen@0.4\dimexpr\macc@kerna
      \let\final@kern#2%
      \ifdim\dimen@<\z@ \let\final@kern1\fi
      \if\final@kern1 \kern-\dimen@\fi
    \else
      \overline{\rel@kern{-0.6}\kern\dimen@#1}%
    \fi
  }%
  \macc@depth\@ne
  \let\math@bgroup\@empty \let\math@egroup\macc@set@skewchar
  \mathsurround\z@ \frozen@everymath{\mathgroup\macc@group\relax}%
  \macc@set@skewchar\relax
  \let\mathaccentV\macc@nested@a
  \if#31
    \macc@nested@a\relax111{#1}%
  \else
    \def\gobble@till@marker##1\endmarker{}%
    \futurelet\first@char\gobble@till@marker#1\endmarker
    \ifcat\noexpand\first@char A\else
      \def\first@char{}%
    \fi
    \macc@nested@a\relax111{\first@char}%
  \fi
  \endgroup
}
\makeatother

\begin{document}

\title{Edge wave and boundary layer of vortex matter}

\author{A. Bogatskiy}
 \affiliation{Kadanoff Center for Theoretical Physics, University of Chicago, 5620 South Ellis Ave, Chicago, IL 60637, USA}
 \author{ P. Wiegmann}
  \altaffiliation{also at IITP RAS, Moscow 127994, Russian Federation}
  \affiliation{Kadanoff Center for Theoretical Physics, University of Chicago,
5620 South Ellis Ave, Chicago, IL 60637, USA}

\date{\today}

\begin{abstract}
We show that a vortex matter, that is a dense assembly of vortices in an incompressible two-dimensional flow, such as a fast rotating superfluid or turbulent flows with sign-like eddies, exhibits (i) a boundary layer of vorticity (vorticity layer), and (ii) a nonlinear wave localized within the vorticity layer, the edge wave. Both are solely an effect of the topological nature of vortices. Both are lost if  the vortex matter  is approximated as a continuous  vorticity patch. The edge wave is governed  by the integrable Benjamin-Davis-Ono equation exhibiting solitons with a quantized total vorticity. Quantized solitons reveal the topological nature of the vortices through their dynamics. The edge wave and the vorticity layer are due to {\it odd viscosity} of the vortex matter. We also identify the dynamics with the action of the Virasoro-Bott group of diffeomorphisms of the circle, where odd viscosity parametrizes the central extension. Our edge wave is a hydrodynamic analog of the edge states of the fractional quantum Hall effect. 
 
\end{abstract}


%47.32.C-       Vortex dynamics (fluid flow)
%05.45.-a       Dynamical systems -- nonlinear
%02.30.Ik       Integrable systems
%47.35.Fg       Solitons -- fluids
%05.45.Yv       Solitons -- nonlinear dynamics of 
%47.27.nb       Boundary layers -- turbulence
%73.43.-f       Quantum Hall effect

\maketitle

\noindent{\bf Introduction} 
In this paper we focus on an exemplary problem of classical hydrodynamics: a blob of vorticity consisting of a dense assembly of sign-like point vortices in 2D incompressible inviscid
fluid, the {\it vortex matter}, or a {\it chiral flow}, see Fig.~1. The question we ask  is: whether and how a `quantization', or discreteness, of vortices affects large scale flows?  

The standard examples of vortex matter are  quantum fluids: rotating superfluid Helium \cite{Khalatnikovbook}, and  optically trapped BEC, see \cite{BEC} and references therein, where  the circulation of  vortices are quantized in units of the Planck constant.  In classical fluids the vortex matter arises in the inverse cascade of a confined turbulent flow when small eddies congregate into large sign-like Onsager's vortex clusters \cite{Onsager,Montgomery}. There are numerous examples in atmospheric, oceanic, aeronautic and astro-physics (tornadoes,  hurricanes, pulsars). The vortex matter is also a subject of what is called quantum turbulence, see \cite{Tsubota2017} for a review. A somewhat related topic is active rotor media,  where fluid particles possess rotational degrees of freedom, see \cite{Irvine}  and references therein. An understanding of the motion of vortex matter is also important for the `vortex method' in computational hydrodynamics, where continuous vorticity is approximated by discrete vortices \cite{Cottet,Majda}. The fundamental importance of the vortex matter had  recently emerged in the theory of fractional quantum Hall effect (FQHE) \cite{W13}. There, electrons  effectively bound to localized magnetic fluxes  move like vortices in  incompressible fluid.

In the listed cases, classical and quantum alike, the vortex matter is a liquid.  However, it is a special class of liquid whose constituents possess a topological characterization, the circulation. Interactions between topological textures, is essentially non-local and have a geometric nature. This makes their flows different.  
The naive coarse-grained approximation where the vortex matter is treated as a uniform vorticity is a classical setting in hydrodynamics, known as a  vorticity patch, or a  Rankine vortex,  a domain of  a uniform vorticity, surrounded by an irrotational flow.  We will show that  this approximation fails: the topological characterization of micro-scale fluid constituents has nowhere to hide. It is observable in hydrodynamics. 

The effects related to the discreteness of the vortex matter are especially pronounced at the boundary. Recall that incompressible fluid with a smooth vorticity and without external forces, or imposed flows  does not allow linear traveling waves.
The same is true for the boundary motion of the Rankine vortex. This changes if the vorticity patch is an aggregation of small vortices, a sort of {\it discretized Rankine vortex}. We will show that the indestructible discreteness of vortices yields a linear {\it edge mode} propagating along the boundary within a new kind of a boundary layer, the  {\it vorticity layer}.  The layer and the edge mode would have been lost had the vortex matter been treated as a continuous vorticity. The vorticity layer acts towards the stabilization of  violent  filaments  in the unstable dynamics of Rankine vortex (see, \cite{Majda,Dritschel} and references therein).
We find the dispersion of the edge mode to be   
\begin{align}
    E_k=\wbar{ U} k+2\eta k|k|,\quad \eta=\Gamma/{8\pi}, \la{109}
\end{align} 
 where  $-\Gamma$ is the circulation of each vortex.
The wave travels against the overall rotation of the layer with velocity \(\wbar U=\Gamma/\sqrt{16\pi l}\), where $l$ is the mean inter-vortex distance. The coefficient \(\eta\) of the non-analytic term in \eq{109} has the dimension of viscosity. We will identify it with the {\it odd viscosity} of \cite{Avron, AW2014}. Our edge mode is a classical prototype of the FQHE edge state \cite{W11}.

 Our main result is the dynamics of the edge mode beyond the linear approximation.  We show that the edge is governed by the celebrated {\it Benjamin-Davis-Ono} (BDO) equation (often abbreviated as Benjamin-Ono)
\cite{BDO}. It is an integrable equation  and exhibits solitons. A remarkable property of BDO is that the solitons possess a quantized vorticity, revealing the topological character of vortices. 

Also, we emphasize another two  results: (i) we identify the boundary conditions at the vorticity jump and the front of ambient irrotational fluid and (ii) establish the relation between the edge dynamics and the action of of the Virasoro algebra, whose central extension happens to be the odd viscosity.

 %%%%%%%%%%%%
 \smallskip
 \noindent{\bf   Kirchhoff equations}  We recall the notion of the Rankine vortex, the naive coarse-grained approximation of the vortex matter. It is a domain \(D\) of uniform vorticity, surrounded by an irrotational flow. The dynamics of the Rankine vortex are the evolution of its boundary, also called  the {\it vorticity jump}. It is governed by the kinematic boundary condition (KBC). The KBC states that the velocity of a fluid  parcel at the interface (the front) equals the velocity of the ambient flow. We denote the vorticity of a clockwise vortex (anticyclonic) patch by \(-2\Omega<0\) and use the frame rotating anticlockwise with the frequency \(\Omega\). Then, if \(z(t)\) is the complex coordinate of a fluid  parcel at the front and \(\uu=u_x-\ii u_y\) is the complex velocity of the flow, the KBC reads
\begin{align}\la{0}
        \dot\bz=\uu|_{z(t)},\quad \uu= -\ii\Omega\bz+
        \ii\frac\Omega\pi\int_{D(t)}\frac{\dd
V'}{z-z'}.
\end{align}
This equation, called {\it contour dynamics} (or CDE), has been extensively studied (see e.g., \cite{Dritschel}).

We will formulate the problem of the `discretized'  Rankine vortex with the help of the Kirchhoff equations (see e.g. \cite{Saffman}). Recall that the complex velocity of the form
\begin{align}\la{3}
        \uu(z,t)=-\ii\Omega\wbar z+
        \frac\ii{2\pi} \sum^N_{i=1}\frac{\Gamma_i}{z-z_i(t)},
\end{align} 
is a solution of  the Euler equation     
\begin{align}\la{1}
        \dot{\bm u}+(\bm u\cdot\nabla)\bm u=-\bm \nabla p
         -2\Omega\times\bm u,
         \quad \nabla\cdot{\bm u}=0  
\end{align} 
%{\red(the Coriolis force is \(-2(\Omega\times\bm u)=2\Omega (-u_y,u_x))\)},
iff the trajectories of the vortices \(z_i(t)\) obey the Kirchhoff equations: 
\begin{align}\label{2}
    \dot{\bar{ z}}_i\equiv \vv_i=-\ii\Omega\wbar {z}_i+\frac\ii{2\pi}\sum^N_{i\neq
j}\frac{\Gamma_j}{z_i(t)-z_j(t)}.
\end{align} 
Both Euler and Kirchhoff equations are written in the rotating frame.
 If the circulations of all vortices  
   are set equal (they do not change due to the Kelvin theorem)
\be
    -\Gamma_i=-\Gamma<0,
\ee
which is the case we consider, the Kirchhoff equations are the discrete version of the contour dynamics \eq{0}. The question we ask is whether the CDE \eq{0} correctly captures the hydrodynamics of the vortex matter. We show that it does not. Even at vanishing spacing between vortices the behavior of tightly packed vortices described by the Kirchhoff equations does not match its naive continuous version.

The naive coarse-grained approximation fails at the boundary. The bulk vortices can indeed be approximated by a uniform  density \(\rho_\infty=2\Omega/\Gamma\) \cite{Feynman}. However, we will see that forces acting within the vortex matter  squeeze the blob  pushing  the boundary  inward. This yields   a singular boundary layer with a sharp peak of vorticity, the {\it overshoot}, as schematically shown on the Fig.~1. Unlike the known boundary layers, such as the Stokes and  Ekman layers, our layer occurs in inviscid fluids. 
The  waves in the overshoot are the edge waves we study. 

To isolate  the effect we focus on the near-circular patch, whose continuous version is a stationary circular Rankine vortex.  In this limit the vortex density \(\rho(\bm r)=\sum_{i\leq N}\delta(\bm r -\bm r_i)\), or vorticity $\omega=-\Gamma\rho$ is approximated by a step function \(\rho_0(r)=\rho_\infty\,{\Theta(R_0\!-\!r)}\)  supported by a disk with the   area  \(\pi R_0^2=N\rho_\infty^{-1}\).   It  shows no dynamics whatsoever: if $D$ is a disk, the r.h.s.\ of \eq{0} vanishes. 

Near-circular, near-stationary  flows are selected by the vanishing  of the angular impulse  (in units of fluid density) \cite{Saffman}  \be {\rm L}\equiv\sum_i \bm r_i\times\bm v_i=0\la{LL}
\ee
(\(\bm v_i\) is the velocity of a vortex), a conserved quantity closely related to the fluid's  angular momentum. The true vortex matter possesses a scale, the inter-vortex {\it spacing} \(l=\rho_\infty^{-1/2}\).

\begin{figure}[h!]
\centering
\scalebox{1.2}{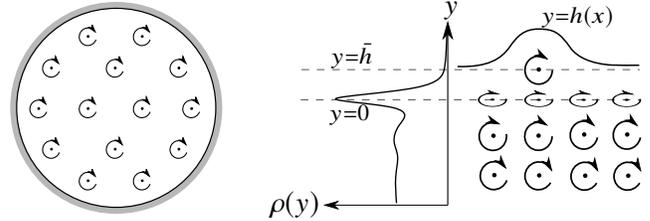} 
\caption{\label{Fig1}Left: the circular blob of vortex matter with a boundary layer. Right: The boundary layer: the blob \(y<0\) is squeezed relative to the continuous vorticity patch \(y<\bar h\); a vortex trapped in the boundary layer $B: 0<y\leq h$ illustrates charge 1-soliton; illustration of the overshoot in the vortex density is on the side.}
\end{figure}

\noindent{\bf Vorticity layer: Overshoot}
The vorticity jump (the boundary of the uniform vorticity domain) of the  Rankine vortex is also the front of the ambient irrotational flow. This is no longer the case if the vortex spacing, no matter how small, is taken into account.

The vortex density significantly departs from the step function \(\check \rho({\bm r})=\rho({\bm r})-\rho_0( r)\) within a few spacing units from the boundary of the equivalent Rankine vortex, forming a boundary singularity, the overshoot, Fig.~1. Numerical data of the stationary overshoot are available from the studies of the electronic density in FQHE (see, e.g., \cite{Morf}), a closely related problem. The data show that the overshoot is an asymmetric peak centered slightly inward, with density oscillations decaying into the bulk. 

We outline the properties of the overshoot  first, before presenting supportive arguments. Some  of them were analytically obtained in \cite{WZ,Can}, others are obtained  here.
We will see that forces acting within the vortex matter squeeze the blob, pushing the vorticity jump inward by the distance \(\bar h=l/\sqrt{8\pi}\). This value agrees with the fitted numerical data (see literature comments). We denote its radial position by \(R\equiv R_0-\bar h\). The squeezing is a new effect (see, however,  literature comments for a related effect in a rotating superfluid confined by a container).
%\cite{Note0}. 

 Inside the squeezed blob $r<R$ vorticity may be treated as uniform \(\rho=\rho_\infty\), and for this reason the circular  vorticity jump $r=R$ does not evolve. In a weakly nonlinear, long-wave approximation that we adopt, we neglect the curvature of the blob and treat the squeezed blob as the lower half-plane \(y<0\) with local coordinates  \(y=r-R,\    \dd x=-R\dd\theta\) relative to the actual vorticity jump. 
  
The gap that emerges between the vorticity jump and the irrotational flow is the boundary vorticity layer. It is a strip with a fixed bottom and a varying top. The top of the layer is the front of the ambient flow, Fig.~1. We denote its graph by \(y=h(x)\):
\be
    B=\{y:\ 0<y\leq h(x)\}
\ee
The front oscillates about the boundary of the equivalent Rankine vortex \(y=\bar h\equiv (1/2\pi)\oint h\,\dd \theta\). 
We  assume that near-stationary flows are in the state of {\it local equilibrium}: the mean radial line density is uniform along the boundary
\be
    \frac\delta{\delta h}
    \ \int \rho(x,y)\,\dd y=\rho_\infty.
    % \int_{y<h(x)}\left[\rho(x,y)-\rho_\infty\right ]\,\dd y=0
    \la{LE}
\ee
The first two moments of the overshoot \(\check \rho({\bm r})=\rho({\bm r})-\rho_0( r)\)  are sufficient to describe  the edge dynamics. They are the line density  and the dipole moment (about the vorticity jump): 
\begin{align}
\begin{split}
     n(x)= \int\,\check\rho(x,y)\,\dd y\la{1208},\quad
     d(x)=- \int y\,\check\rho(x,y)\, \dd y.
\end{split}
\end{align}
The former is  the line density of vortices trapped in \(B\): $n(x)=\int_B\rho(x,y)\,\dd y$. It follows from \eq{LE} that 
\be
    n(x)=\rho_\infty(\bar  h- h(x)).\!\la{1209}
\ee
This is our edge mode.

We will show that the means of the dipole moment \(\bar d=(1/2\pi)\oint d\dd \theta\) and the squeezing \(\bar h \) (which are also their stationary values) have the exact values
\begin{align}
       \bar d= 1/8\pi,\quad \bar h^2=l^2\bar d.
\la{100}
\end{align} 
In the limit of the vanishing spacing \(l\to 0\), the dipole moment \(\bar d\) does not vanish. This means that the density possesses a singular double layer at the vorticity jump \cite{WZ,Can}. 
Then we can  restate (\ref{1208}-\ref{100}) as a singular expansion about  the actual vorticity jump at $y=0$ 
\begin{align} \la{1109}
\begin{split}
&\check\rho(x,y)=-n(x)\,\delta(y)+ d (x)\,\delta{'}(y)+\dots\\
 & \rho(x,y)\!=\!\rho_\infty\left(\Theta(-y)\!+\!
    h\delta(y)\!-\!\frac {\bar h^2}2\delta{'}(y)\right)+ d \delta{'}(y)\!+\!\dots 
   \end{split}
    \end{align}
     obtained  by expending  the step function \(\rho_0(y)\equiv\rho_\infty\Theta(\bar h-y)\) about $y=0$. 
The terms in \eq{1109} are: the squeezed blob of the uniform vorticity, the simple, and the double layers.
%(the density in \(B\) is restricted by  (\ref{1208}-\ref{1209}) equivalent to $n(x)=\int_B \rho dy,\;\int_B %y\rho dy=0$.
The simple layer indicates that the azimuthal velocity jumps
\begin{align}
    U (x)\equiv(1/2)\underset{y=0}{\rm disc}\;[u_x]=\Omega h(x). \la{808}
\end{align}
The layer speeds up by the mean velocity \(\wbar{U}=\Omega\bar h\) with respect to the squeezed blob. 

We will also see that the  moments  are not independent. A balance of forces at the vorticity jump requires that the height \(h\) and the moment \(d\) move subject to the relation 
\begin{align}
   d-  hd^H_x=\left(h/l\right)^2,
  \la{41}
\end{align}
where  
\be
    f^H=\frac 1{\pi}\int \frac{f(x')-f(x)}{x'-x}\dd x'
\ee
is the Hilbert transform, and \(d_x=\p_xd\). Relation \eq{41} also means that the dipole moment
of the density $\bar d=\int (h-y)\rho dy$ about the moving front $y=h(x)$ is a constant equal to \(1/(8\pi) \) in  agreement  with \cite{WZ}.

Now we present some arguments in support of these claims.

\noindent{\bf The angular impulse}  
The value of the mean dipole moment \eq{100} promptly follows from the identity
\begin{align}
        {\rm L}=\Omega \sum r_i^2-( \Gamma/4\pi)N(N-1).\la{13}
\end{align}
It follows  after a multiplication  of the Kirchhoff equations \eq{2} (with all  $\Gamma_i$ equal) by   \( z_i \) and summation  over all vortices. 
The last term is the number of vortex pairs. The term \((\Gamma/4\pi) N^2=\Omega\int\rho_0
r^2 \dd V\) is the  angular momentum  of the continuous Rankine vortex. The sub-leading
term \(( \Gamma/4\pi)N\) counts the excluded terms \(i=j\) in the sum \eq{2},  reflecting the discreteness of the vortex matter. Hence
\begin{align}\la{101}
        {\rm L}=\Omega \int r^2 \check\rho \, \dd V+(\Gamma/4\pi) N.
\end{align}
Then  \eq{1208} implies 
\be
    {\rm L}=N\Gamma(-2\bar d+1/4\pi).
\ee 
We conclude that  \({\rm L}=0\) flows possess a dipole moment  equal to \(\bar d=1/8\pi\) as in  \eq{100}.

\noindent{\bf Stream function} A more detailed structure of the overshoot follows from  the evaluation of   the stream function
\begin{align}
    \Psi({\bm r})=(\Gamma/{2\pi})\int\log|{\bm r}-{\bm r}'|\rho\,\dd V'-(\Omega/2) r^2
\end{align}
 and velocity \(u_x=\p_y\Psi,\ u_y=-\p_x\Psi\).  Eqs.\ (\ref{1208}-\ref{1209}) suffice to determine them outside but near $B$.
%expansion of the Newton potential around %\(r=R\)
%\begin{align*} 
%    \log&|{\bm r}-{\bm r}'|=\log|x-x'|+\\ %&|{y-y'}|\left(\pi\delta_ %{xx'}+R^{-1}\Theta(y-y')\right)+\frac{1}{2}\p_x%\frac{{(y-y')}^2}{x'-x}+\dots
%\end{align*}
%Applying it to  the singular expansion of the %density \eq{1109}, and taking into account %(\ref{1208}-\ref{1209}), we obtain the %expansion of the stream function outside of %\(B\). 
The stream function  consists of the rotational part representing the solid rotation of the squeezed blob
\begin{align}
y\notin B:\  \Psi(x,y)=\Omega(2\bar h y- y^2)\Theta(y)+\psi(x,y) \la{2309}
\end{align}
and the irrotational part generated by the double layer
\begin{align}
    \psi\!=\!\frac{\Gamma}{2\pi}\p_y\!\oint \!\log|z\!-\!x'| d(x') \dd x'\!\approx\!\frac\Gamma{2}\left(\sign (y) d+yd^H_x\right)\la{psi}
\end{align}
Contrary to the Rankine vortex the stream function jumps through  the double layer.
These formulas and \eq{808} yield  the boundary values of the velocity 
\begin{align}
y=\pm 0:\quad  &u_x=2U\Theta(y)+(\Gamma/2)d^H_x, \   u_y=\pm(\Gamma/2) d_x,\la{09011}\\
%&\psi=\Omega(2\bar h y- y^2)\Theta(y)+\frac\Gamma{2}\left(\sign (y) d+yd^H_x\right) ,\la{2309}\\
%&
y=h:\quad &u_x=\Gamma  n+(\Gamma/2)d^H_x, \qquad\  u_y=-(\Gamma/2)d_x. \la{0901}
\end{align}
For reference we express the boundary values of the stream function through the height of the front assuming relation  \eq{41} 
\begin{align}
&\restr\Psi {\mp 0}=\mp(\Gamma/ 2) d=\mp\left(\Omega h^2-\eta h_x^H\right),\la{2411}\\
    &\restr\Psi{h}=
    \Omega(2\bar h h-h^2)+\frac\Gamma{2}\left( d+hd^H_x\right)=
    2\Omega\bar h h+ 2\eta h^H_x,\la{241}
    \end{align}
where \be\eta=\Gamma \bar d\la{eta}\ee  is the mean dipole moment of vorticity.
%%%%%%%%%%%%

\noindent{\bf  Balance of forces } Now we are ready to obtain \eq{41}.  It  follows from the principle of {\it continuity} commonly used in weak solutions of hydrodynamics.

We look for the extremum  of the free energy   
\begin{align}
   {\rm E}-2\Omega{\rm L}=\left[\int_{r<R}\left(\frac{{\bm u}^2} 2-2\Omega\Psi\right)+\int_{r>R}\frac{{\bm u}^2} 2\right]\dd V.
\end{align}  
%(the singular point of the vorticity jump  %$r=R$ should be excluded from the integration). 
with respect to the squeezing length $h=R_0-R$. 
The extremum condition $\delta ({\rm E}-2\Omega{\rm L})=0$  is equivalent to  the continuity of the free energy density across the jump. It ensures the balance of the Coriolis force acting on the vorticity jump from the bulk side and the centrifugal force caused by the squeezing
\begin{align}
\text{balance at}\ y=0:\  -2\Omega\restr\Psi{-0}=(1/2)  \underset{y=0}{\rm disc}\;[u_x^2].\la{27}
\end{align}
%\begin{align}
%        & y<-\bar h:\quad\dot{\check \phi}+ p+\12{ \check u}^2=2\Omega\check\psi,\la{262}\\
%        & y>-\bar h:\quad \dot{\check \phi}+ p+\12{ (U+\check u)}^2=-2\Omega\check\psi.\la{27}
%\end{align}
The boundary values of the stream function  and the velocity entering the balance equation are given by  (\ref{2411},\ref{09011}). Plugging them in we obtain
\begin{align}\la{1908}
   \Omega  \Gamma d=2U^2+\Gamma U d_x^H.
\end{align}
This is relation \eq{41}.

% We illustrate the balance condition by integrating the Euler
% equation over  a box of vanishing width surrounding  a line element of the
% jump. Consider,  \(y\)-component \(\int^\e_{-\e}\dot u_ydy+\int_{+0}^\e\omega u_xdy+2\Omega\psi|_{-0}+{\rm disc[}\12 u^2+p]=0\).  Exclusion of the divergent part at \(y=0\) in the integral version of the Euler equation is the physical assumption. The balance relation follows under condition that pressure is continuous  and \(\omega u_y\) taken values in \(B\),  and \(\dot u_y\) are finite.  

\noindent{\bf Kinematic boundary conditions}  Having proven  relation \eq{41} and expressed the stream function in terms of the height of the front,  we now employ the KBC to produce the edge dynamics. In the rotating frame the KBC reads
\be
        \text{KBC:}\qquad \left(\p_t-U\p_x\right) h=-\frac \dd{\dd x}[\restr\Psi{h(x)}].
\la{141}
\ee
Combining the KBC with \eq{241} we obtain the dynamics of the front, our main result.

\noindent{\bf Main result: Benjamin-Davis-Ono equation} 
The number of vortices \eq{1209} trapped in the line element of \(B\) evolves according to the BDO equation \cite{BDO}
\be\label{44}
      \text{BDO}:\   (\p_t+\wbar{U}\p_x)n-(\Gamma/ 2)\p_ x \left( n^2 -4{\bar d} n^H_x\right)=0.
\ee
The  linearized BDO admits linear modes with dispersion \eq{109}. The BDO is an integrable equation with  explicit periodic and solitonic solutions \cite{Matsuno}. A remarkable property distinguishes BDO from other solitonic equations: the first integral of BDO, the `charge' is {\it quantized} in units of \(8\pi\bar d\) ($2\pi$ times the ratio of coefficients before the dispersive and the non-linear terms). For  \(\bar d\) given by \eq{100} the soliton charge is integer
\begin{align}
        \text{quantization: }\; \oint n \,\dd x\in \mathbb Z .
\end{align}
The quantization of solitons is our major result: the edge mode is a motion of discrete vortices trapped in the overshoot. Naturally, their number is integer. Explicitly, the 1-soliton solution in the frame co-moving with the boundary layer, \(\xi=x-\wbar{U} t\), is 
\begin{align}
        \text{soliton}: \  n_s(\xi,t)=  \frac {8\bar d A}{(\xi+v_st)^2+A^2},\quad
v_s=\eta{ A}^{-1}.\nn  
\end{align}
It carries precisely one vorticity quantum. The constant \(|A|\gg l\) determines the width, the height, and the speed of the soliton.  Bumps \(A>0\)  travel faster, \(v_s>0 \), than the overall rotation of the layer, dents   \(A<0\) travel slower, \(v_s<0\).

In the remaining part of the paper we obtain the boundary conditions for the vortex matter and  identify the coefficient \(\eta\) \eq{eta}  with the odd viscosity.

\noindent{\bf   Bernoulli equation and pressures}  The  vorticity of the  flow  outside the boundary layer is uniform: $\omega=-2\Omega\Theta(-y)$.  Such   flows are governed by the Bernoulli equation. Let us introduce the hydrodynamic potential  \(\varphi\) of the irrotational flow generated by the double layer, the harmonic conjugate of $\psi$, whose boundary value is $\varphi=\sign(y)\psi^H=(\Gamma/2)d^H$ given by \eq{psi}. Then 
\begin{align}
    y\notin B: \quad\dot\varphi+ u^2/2+p=0.\la{B}
\end{align}
Since we already know the dynamics of the front, the dynamics of \(\varphi\) can be read off the pressure.  Omitting the algebra we present the leading gradient approximation of the Bernoulli equation evaluated at both boundaries $y=0$ and $y=h(x)$:
\begin{align} \la{29}
     y\in  \p B:\quad \left ({\p_ t}+ U {\p_x}\right)\varphi=2\eta\p_x u_y.%\quad u_y=-\phi^H_x.   
\end{align}
Comparing \eq{29} to \eq{B} we find the pressure at the  boundary 
\begin{align}\la{3109}
     y\in \p B:\quad p+U^2/2=-2\eta\p_xu_y.
\end{align}
The centrifugal energy in \eq{3109} is an effect of squeezing.

%%%%%%%%%%%%%% %%%%%%%%%%%%%%
\noindent{\bf Anomalous stress, odd viscosity and dynamics boundary  conditions}  The  boundary moves to offset the stress the vortex matter exerts on the ambient fluid. Eq.\ \eq{29} already gives the trace of the stress at the boundary since 
 \(
    2p+U^2=-(\tau_{xx}+\tau_{yy}).
\) 
It can be harmonically extended toward the bulk
\be\la{3702}
\tau_{xx}+\tau_{yy}=4\eta\p_x u_y.
\ee
The traceless part of the stress follows  from the  requirements that the  stress be dissipation-free, describes an isotropic fluid and is linear in velocity. The stress meeting these requirements has been introduced in \cite{Avron} (independently from vortex matter). It is a spin-2 tensor
\begin{align}
        \la{320}
        \tau_{xx}- \tau_{yy}= 2\eta\,(\p_x u_y+\p_y u_x),\
        \tau_{xy}= \eta\,(\p_y u_y-\p_x  u_x).
\end{align}
The coefficient $\eta$ was called odd viscosity.
 
Later it was found \cite{AW2014} that the boundless vortex matter features the same stress (it  was called  there the {\it anomalous stress}). The coefficient $\eta$ computed in   \cite{AW2014}  was found to be a universal fraction of the circulation quantum \(\eta=\Gamma/8\pi\) identical to \eq{eta}. We now see that  the odd viscosity has a new interpretation as the dipole moment of the overshoot \eq{eta}.

 Comparing with \eq{320} we observe that the normal component of the stress vanishes. Hence, on the boundary \begin{align}\la{1709}
  \text{DBC}:\quad    \restr{\tau_{yy}}{\p B}=0.
\end{align}
 This could be used as a boundary condition for the vortex matter which together with  \eq{3702} determines the stress tensor and the flow outside of the boundary layer:
 the trace of the stress (the pressure)  is continuous through the boundary layer, the normal stress vanishes, the shear and the tangential components are related by the 
 Cauchy-Riemann condition $\tau_{xx}=2\tau_{xy}^H$. Often the boundary value of the stress 
 is called Dynamic Boundary Condition (DBC). Notice that our DBC differs from the stress-free conditions of a free surface, but is similar to that of a surface covered by an inextensible film.
 
 We conclude by formulating the bulk dynamics in the linear approximation
 when we replace $U$ by $\wbar{U}$ in \eq{1709}. We introduce the holomorphic  hydrodynamic potential \(\phi=\varphi+\ii\psi\) ($\uu=\p\phi,\ \bar\p\phi=0,\ 2\p=\p_x-\ii\p_y$) and extend \eq{29} toward the bulk of the blob. The result is 
\begin{align} \la{2912}
     y<0:\quad \left ({\p_ t}+ \wbar{U} {\p}\right)\phi=\tau, \quad \tau=2\ii\eta\p^2\phi,
\end{align}
 where $\tau=(1/2)\left(\tau_{xx}-\tau_{yy}\right)-\ii\tau_{xy}$  is the holomorphic component  of the stress \eq{320}.  
 
 \smallskip
\noindent{\bf  Edge dynamics as  action of the   Virasoro-Bott group with odd viscosity as a central extension} The edge dynamics has a natural connection to the Virasoro-Bott group: the centrally extended orientation preserving diffeomorphisms of a circle (see, e.g., \cite{Vir} on  Virasoro-Bott group and hydrodynamics). We  briefly comment on this point.  Let us pass to the Lagrangian specification of the  irrotational flow outside the blob.  We map conformally the outer domain $y>h(x)$  to the upper half plane ${\rm Im} \,z>0$, evaluate the dynamics of the hydrodynamic potential at the front $\Phi=\restr\phi{h}$ and extend it holomorphically  to ${\rm Im} \,z>0$. With the help of the transformation \(\dd \Phi =\restr{\left(\dot\phi  +u_y\dot h\right)}{h}\dd t =\restr{\left(\dot\phi +u_y^2\right)}{h}\dd  t\),  we obtain 
\begin{align}\la{2911}
    {\rm Im} \,z=0:\quad &\left(\p_t+ U \p_z\right)  \Phi= T_{zz},
\end{align} 
Here $T_{zz}$  is the boundary value of the holomorphic tensor \be\la{T_{zz}}
     T_{zz}=-(\p_z\Phi)^2+2\ii\eta
    \p_z^2\Phi.\ee 
Eq.\ \eq{2911} is a form of the BDO equation \eq{44} that is readily analytically continued outside of the blob. It underlines the geometric meaning of the edge dynamics. The  dynamics is merely an action of the Virasoro-Bott group. $T_{zz}$ is the stress-energy tensor of Conformal Field Theory, a generator of the Virasoro Poisson-Lie algebra. Omitting the calculation
%With the help of the   Poisson brackets  $\{h(x),\ h(x')\}=   %\Omega^{-1} \delta_{xx'}^{'}$ and the relation %$\p_z\phi|_h\approx(\Gamma/2)\p_x(d^H+\ii d)$ followed from %\eq{0901} 
we present the  brackets for the  generators  $L_n=(1/2\pi)\oint z^{n-1} T_{zz}\dd z$ of the Virasoro algebra
%$\{\uu({\bm r}),\uu({\bm r}')\}=\Omega\ii\delta_{{\bm r}{\bm r}'}$  
\begin{align}
%(-\Omega)^{-1}\{T(x),\ T(x')\}=2\left(T( x)+T(x')\right)\delta_{xx'}'-\eta^2 \delta^{'''}_{xx'}.
(\pi/16\Omega\ii)\{L_n,\ L_m\}=(n-m)L_{n+m}+\eta^2 n^3\delta_{n+m,0},\nn
\end{align}
whose central extension in this case is set by the odd viscosity.
\smallskip

\noindent{\bf Summary} Summing up, we demonstrated that the vortex matter cannot be approximated by a patch of uniform vorticity. The discreteness of vortex matter is revealed on the edge in the form of edge waves and quantized solitons. The edge wave propagates along a new kind of boundary layer, the overshoot of vorticity.  Contrary to the known boundary layers the vorticity layer offsets the dissipation-free stress of the vortex matter, known as odd viscosity stress. We identify the odd viscosity kinetic coefficient with the dipole moment of the vorticity overshoot and formulate specific boundary conditions for the vortex matter.  We also underline the geometric nature of vortex matter by identifying its flows with the action of the Virasoro-Bott group and the odd-viscosity with the central extension.

Apart from fluid mechanics our results may be relevant for physics of rotating superfluid, optically trapped atomic gases, and especially for FQHE, since our edge mode is a classical prototype of the FQHE electronic edge state. 

It is tempting to think that, in general, flows of topological matter reveal their quantized topological numbers through the non-linear edge dynamics.

\noindent{\bf Literature comments} The idea to treat vortices as a continuum medium goes back to Onsager's 1949 work \cite{Onsager}; Flows of the vortex matter  are the standard subject of the theory of rotating  superfluids \cite{Khalatnikovbook}; In independent 1960 papers  Hall \cite{Hall} and Kemoklidze and Khalatnikov  \cite{KK} argued that vortices in rotating superfluid Helium repel from  walls of the container due to the  effect of excluded volume. This effect is the bulk precursor of the squeezing effect; The overshoot was seen numerically in the FQHE setting in \cite{Morf}. The numerical value \eq{100} of  squeezing \eq{1209} has been noticed by A.G. Abanov (unpublished) by examining the numerical data of A. Shitov (unpublished); The value of the dipole moment has been obtained in \cite{WZ} and further analyzed in \cite{W11, Can}; The BDO equation appears in the description of various fluid  interfaces: density \cite{BDO}, vorticity \cite{Stern}, or shear \cite{Maslowe}. More recently it emerged  as a hydrodynamic description of the Calogero model \cite{ABW}, and as a theory of edge states in the fractional quantum Hall effect \cite{W11};  In the interesting paper \cite{B} Bettelheim showed that the hydrodynamics of the Calogero model can be seen as a 2D incompressible hydrodynamics in the phase space. This work also emphasized the relation between the actions of the area preserving diffeomorphisms and the Virasoro algebra at the boundary;  Motivated by the adiabatic transport in moduli space computed by Avron, Seiler and Zograf \cite{ASZ} for quantum Hall effect, Avron \cite{Avron} discussed a model of compressible hydrodynamics with odd viscosity. There the odd viscosity was introduced phenomenologically, and with no connection to the vortex matter. A comprehensive list of references on recent developments can be found in the paper \cite{ACG}, which studies the effects of the anomalous stress on free surface waves. Despite some similarities, there is an important difference with our study. In  \cite{ACG} the anomalous stress \eq{320}  were phenomenologically imposed on a flow  with zero net vorticity. In the vortex matter  the anomalous stress and the net vorticity can not be considered separately; Finally we mention the recent work on assemblies of active rotors, see \cite{Irvine} and references therein. There again, despite some similarities, the short-range  nature of  interactions and damping make active rotors different from the vortex matter.

We thank  A. G. Abanov, G. Monteiro, S. Ganeshan, W. Irvine, V. Vitelli, S. Llewellyn-Smith, D. Dritschel and E. Sonin for helpful discussions. PW acknowledges support from  the Brazilian Ministry of Education (MEC) and the UFRN-FUNPEC and International Institute of Physics, and  support  from the Simons Center for Geometry and Physics, Stony Brook University during the work on this paper.


\begin{thebibliography}{1}


\bibitem{Khalatnikovbook} I. M. Khalatnikov, \emph{An Introduction to the Theory of Superfluidity}, Westview Press, 2000; E. B. Sonin, \emph {Dynamics of quantised vortices in superfluids,} Cambridge University Press, 2016.
\bibitem{BEC}
 G. Gauthier, M.T. Reeves, et al., \emph{Negative-temperature Onsager Vortex Clusters in a Quantum Fluid}, arXiv:1801.06951, 2018;
\bibitem{Onsager}
L. Onsager, Nuovo Cimento, Suppl. {\bf 6}: 249, 279 (1949), see also review
of Onsager's works by G. L. Eyink, K. R. Sreenivasan, \emph{
Onsager and the theory of hydrodynamic turbulence}, Rev. Mod. Phys,  \textbf{78}:
87 (2006).
\bibitem{Montgomery} D. Montgomery and G. Joyce, \emph{Statistical mechanics of ?negative temperature" states}, Phys. of Fluids 17.6: 1139 (1974).
%  
% \bibitem{BEC}
% M. R. Matthews, B. P. Anderson,  P. C. Haljan,  D. S. Hall,  C. E. Wieman, %and E. A. Cornell,  \emph{Vortices in a Bose-Einstein Condensate}, Phys. %Rev. Lett., \textbf{83}: 2498 (1999).  

 \bibitem{Tsubota2017} For a review see Tsubota M, Fujimoto K, Yui S. \emph{Numerical studies of quantum turbulence}, J.  Low Temp. Phys.  1;188:119 (2017).
%S. P. Johnstone, A. J. Groszek et al., \emph{Order from chaos: %Observation of largescale flow from turbulence in a two-dimensional %superfluid}, arXiv:, arXiv:1801.06952,2018.
\bibitem{Irvine} B. C. van Zuiden, J. Paulose, W. T. M. Irvine, D. Bartolo, V.  Vitelli, {\it\  Spatiotemporal order and emergent edge currents in active spinner materials}. Proc. Natl. Acad. Sci. USA 113, 12919 (2016); 
D. Banerjee, A. Souslov, A. G. Abanov, V. Vitelli, {\it Odd viscosity in chiral active fluids}
Nature Commun. 8, 1573 (2017).
 \bibitem{Cottet} Cottet, G.-H., Petros D. Koumoutsakos,
\emph{Vortex Method: Theory and Practice}, Cambridge University Press, 2000.
\bibitem{Majda} A.~J. Majda, A.~ L. Bertozzi, \emph{Vorticity and Incompressible Flow},  Cambridge University Press, 2002. 
\bibitem{W13}
P.~B. Wiegmann,
\newblock \emph{Hydrodynamics of {E}uler incompressible fluid and the fractional
  quantum {H}all effect}.
\newblock { Phys. Rev. B}, {\bf\ 88}: 241305, (2013).\\
P.~Wiegmann,
 \emph{Anomalous hydrodynamics of fractional quantum {H}all states}.
 { J.of Exp. Theor.Phys.}, \textbf{117}(3):538 (2013).
\bibitem{Dritschel}D.~G. Dritschel, \emph{The repeated filamentation of two-dimensional
vorticity interfaces} J. Fluid Mech, \textbf{ 194},  511 (1988); D. Crowdy, J. Marshall, {\it Analytical solutions for rotating vortex arrays involving multiple vortex patches}, J. Fluid Mech., \textbf{523}, 307-337 (2005).



\bibitem{Avron}J. Avron, \emph{Odd viscosity}, J.  Stat. Phys. {\bf\ 92},
543 (1998).

\bibitem{AW2014}
P.~Wiegmann and A.~G. Abanov,
\newblock \emph{Anomalous hydrodynamics of two-dimensional vortex fluids},
\newblock Phys. Rev. Let. \textbf{113}: 034501 (2014) (there the stress is defined with the opposite sign to \eq{320}).
\bibitem{W11} {Wiegmann, P},
 \newblock \emph {Nonlinear hydrodynamics and fractionally quantized solitons
at the fractional quantum Hall edge},
 \newblock 
         { Phys. Rev. Let.},
         \textbf{108:},
         {206810} (2012).
\bibitem{BDO}  R. E. Davis and  A. Acrivos, \emph{Solitary internal waves
in deep water} J . Fluid Mech., {\bf\  29}: 593, (1967).\\
T. B. Benjamin,
\emph{Internal waves of permanent form in fluids of great depth}  J. Fluid
Mech. {\bf\ 29}: 559 (1967),\\ H. Ono, \emph {Algebraic solitary waves in
stratified fluids},
 J. Phys. Soc. Japan {\bf\ 39}: 1082 (1975).


\bibitem{Feynman}Feynman, R. P. 1955. {\it\ Application of quantum mechanics
to liquid Helium}; p 17 of: Gorter, C. J. (ed.), Progress in Low Temperature
Physics, vol. 1. North-Holland.
\bibitem{Saffman}
P.G. Saffman, \emph{ Vortex dynamics}, Cambridge Univ. Press, 1992.
 %V. V. Kozlov, \emph{General Theory of Vortices}, %Springer,2003.


\bibitem{Morf} N. Datta, R. Morf, R. Ferrari, \emph{Edge of the Laughlin
droplet}, Phys. Rev. {\bf  B 53}, 10906 (1996), see also \cite{Can} for references
on  early  numerical evidence of the overshoot.

\bibitem{WZ} A. Zabrodin and P. Wiegmann, \emph{Large-N expansion for the
2D Dyson gas}, J. Phys. A {\bf 39}: 8933,  (2006).

\bibitem{Can} T. Can, P.J. Forrester, G. T\`ellez and P. Wiegmann, \emph{Singular
behavior at the edge of Laughlin states}  Phys. Rev. {\bf  B} 89, 235137 (2014).

\bibitem{Matsuno}Y. Matsuno, \emph{Bilinear Transformation Method}, in v.
174, Math. Sci.  Eng., Academic Press, 1984.

\bibitem{Vir}  V.I. Arnold, B. A. Khesin, Topological methods in hydrodynamics, Ch. VI, Springer 1999.

\bibitem{Hall} Hall, H. E. "The rotation of liquid helium II." Advances in Physics 9: 89-146 (1960).

\bibitem{KK} M. Kemoklidze and I. Khalatnikov, \emph{ Hydrodynamics of rotating
 Helium II in an annular channel}, JETP {\bf46}:1677 (1961).
 \bibitem{Stern} 
M. E. Stern and  L. J. Pratt, \emph{Dynamics of vorticity fronts}, J.Fluid Mech.  {\bf\ 161}:  513 (1985). 
\bibitem{Maslowe}S. Maslowe and  L. Redekopp, \emph{Long non-linear waves
in stratified shear flows}, J.Fluid Mech. {\bf\ 101}: 321 (1980).
\bibitem{ABW} A.G. Abanov,  and P. B. Wiegmann,  \emph{Quantum hydrodynamics, the quantum Benjamin-Ono equation, and the Calogero model}, Phys. Rev. Lett. {\bf 95}: 076402 (2005).M. Stone, A.  Inaki, and X. Lei, \emph{The classical hydrodynamics of the Calogero-Sutherland model} J. Phys. A {\bf  41}: 275401 (2008); A. G., Abanov, E. Bettelheim,  P.  Wiegmann,\emph{
Integrable hydrodynamics of Calogero-Sutherland model: bidirectional Benjamin-Ono
equation}. J. Phys. A {\bf  42}: 135201 (2009).

\bibitem{B} E. Bettelheim,  \emph{Integrable quantum hydrodynamics in two-dimensional phase space} J. of Phys. A:  {\bf 46}: 505001 (2013).

\bibitem{ASZ} J. Avron, R. Seiler and P. G. Zograf, \emph{Viscosity of quantum
{H}all fluids}, Phys. Rev. Lett. {\bf 75}: 697 (1995).

\bibitem{ACG} A. G. Abanov, T. Can, S. Ganeshan, \emph{Odd surface waves
in two-dimensional incompressible fluids}, SciPost Physics 5:010 (2018).



%\bibitem{CambellKranov81} L. J. Campbell, Yu K. Krasnov, {\it Edge waves
%&of a vortex continuum}, Phys. Lett. {\bf A} 84:75 (1981):



%\bibitem{Abanov} A.G. Abanov, 2013, unpublished.










% B. A. Kupershmidt, P.  Mathieu, {\it  Quantum KdV like equations and perturbed Conformal Field
%theories}  Phys. Lett. B227, 245 (1989) 
%B. Khesin, and G. Misiolek. "Euler equations on homogeneous spaces and Virasoro orbits." Advances %in Mathematics 176.1,  116 (2003)

%\bibitem{Note0} An analog 
%of this effect  is known in the theory of superfluids since 1960's, see
%literature comments.

%%\bibitem{Note1}The  relation \eq{41} also means that the dipole moment
%%computed with respect to the moving front $y=h(x)$ is a constant equal to
%%\(1/(8\pi) \) in  agreement  with \cite{WZ}.

% \bibitem{Note2}  Writing \eq{1109} we represented  the step  \(\rho_0(y)\equiv\rho_\infty\Theta(\bar h-y)\) in (\ref{1208})  as a singular expansion about $y=0$: \(\Theta(\bar h-y)=\Theta(-y)+\bar h\delta(y)-\12
% \bar h^2\delta'(y)+\dots\).
\end{thebibliography}
\end{document}https://www.overleaf.com/5725199413